\documentclass[aps,pra,twocolumn,showpacs,floatfix]{revtex4}

\usepackage[pdfview=FitV,pdfstartview=FitV]{hyperref}   
\usepackage{amssymb}    
\usepackage{graphicx}   
\usepackage{epsfig}
\usepackage{verbatim}   
\usepackage{color}      

\newcommand{\bfr}{{\mathbf{r}}}

\newcommand{\etal}{{\textit{et al.},}}

\begin{document}

\title{Quantum Simulators at Negative Absolute Temperatures}

\author{\'Akos Rapp}
\affiliation{ Institut f\"ur Theoretische Physik, Leibniz Universit\"at, 30167 Hannover, Germany}
\affiliation{ Institut f\"ur Theoretische Physik, Universit\"at zu K\"oln, 50937 K\"oln, Germany}

\date{\today}

\begin{abstract}
We propose that negative absolute temperatures in ultracold atomic clouds in optical lattices can be used to simulate quantum systems in new regions of phase diagrams. First we discuss how the attractive SU(3) Hubbard model in three dimensions can be realized using repulsively interacting $^{173}$Yb atoms, then we consider how an antiferromagnetic S=1 spin chain could be simulated using spinor $^{87}$Rb or $^{23}$Na atoms. The general idea to achieve negative absolute temperatures is to reverse the sign of the external harmonic potential. Energy conservation in a deep optical lattice imposes a constraint on the dynamics of the cloud, which can relax toward a $T<0$ state. As the process is strongly non-adiabatic, we estimate the change of the entropy.
\end{abstract}


\maketitle

The aim of dedicated quantum simulations is to study experimentally the properties of a complex quantum system by another quantum system which is less complicated or, at least, better understood. Among these approaches one has to mention graphene, where electrons can be described by a $2+1$-dimensional Dirac equation~\cite{graphene}; bilayer $^3$He, which shows quantum critical behavior~\cite{bilayer3He} that resembles heavy fermion materials~\cite{heavyfermions}; atom corals~\cite{atomcoral} where quantum billiards could be simulated; or cavity quantum electrodynamics where one can study the coupling between light and matter~\cite{cavityQED}. 

Ultracold atomic clouds nevertheless have a distinguished place among quantum simulations. During the last two decades, experimental control developed to a high level in the tuning of microscopic parameters and in detection. Ultracold atomic clouds are also very well isolated and free of unwanted perturbations, like disorder, phonons, etc. Finally, the relevant time and length scales are typically larger than in solid-state systems; therefore the behavior is more tractable experimentally, even in out-of-equilibrium situations.

A number of experiments with ultracold atomic clouds benefited from the advantages discussed above. In addition to the realization of the Hubbard model for two fermionic species~\cite{SU2HUbbard,SU2HubbardETH} (highly relevant for high transition temperature superconductivity~\cite{AndersonHighTC}), ultracold atomic clouds have been used to study the Mott transition for bosons~\cite{BoseHubbard} and the
Anderson localization by a disordered potential~\cite{disorder}, but also more abstract concepts such as gauge fields~\cite{gaugefields,gaugefields-Bloch} and black holes~\cite{blackholes}.

In this work, we shall investigate how negative absolute temperatures, $T<0$, in ultracold atomic experiments can grant access to unexplored regions of phase diagrams. After a general discussion, we will focus on two model Hamiltonians.

The idea of negative absolute temperatures is neither complicated nor new \cite{Landau,Kittel}. The reason why $T<0$ is only seldom discussed is that for most physical systems, negative $T$ is not possible in equilibrium. However, as proposed in Refs.~\cite{Mosk,negT}, ultracold atomic clouds in optical lattices can exhibit $T<0$, presenting new possibilities to study correlated quantum systems.

To gain some insight about the $T<0$ regime, and to show connections to the more familiar case of $T>0$, let us consider the equilibrium partition function of a quantum system,
\begin{equation}
  Z \equiv {\rm Tr} \; {\rm exp}( - \beta H) = \sum_n e^{-\beta E_n}. \label{eq:Z}
\end{equation}
Here $\beta=1/T$ is the inverse temperature (we set the Boltzmann constant $k_B =1$) and $H$ is the Hamiltonian with $n$ and $E_n$ denoting (many-body) eigenstates and the corresponding energies. There is no principle which forbids $\beta < 0$ in general. Nevertheless, most physical systems have a spectrum $E_n$ that is bounded from below but has no upper bound. The simplest example of such a spectrum is the kinetic energy of free particles, $\sim \mathbf{p}^2$, or for ultracold atoms the trapping potential $\approx V_0 \mathbf{r}^2$ with confinement $V_0 > 0$. It is clear from Eq.~(\ref{eq:Z}) that without an upper bound in energy, no equilibrium is possible at $\beta < 0$ as the partition function diverges. Although this is the case for the vast majority of physical systems, it is not impossible to construct Hamiltonians which have only an \emph{upper bound} in energy, where, in turn, only $\beta <0$ is possible in equilibrium. We shall below focus on how this can be achieved with ultracold atoms.

From Eq.~(\ref{eq:Z}) we can also see that while for $\beta > 0$ the lower energy states have higher occupation probabilities $P(E)\sim e^{-\beta E}$, the opposite, an ``inverse population'', is realized at $\beta < 0$. Inverse population of the energy levels leads to an unusual condition in matter, which is most widely used in lasers \cite{lasers}.

The most important consequence of Eq.~(\ref{eq:Z}) is that the phase diagram of $H$ with $\beta < 0$ can be mapped to the phase diagram of a system with $\tilde \beta = -\beta$ described by $\tilde H = - H$. This is the equivalence that could be used in quantum simulations to chart new parts of phase diagrams experimentally, as we shall discuss later in the context of two models.

\begin{figure}
 \centering
 \includegraphics[width=0.45\textwidth,clip=true]{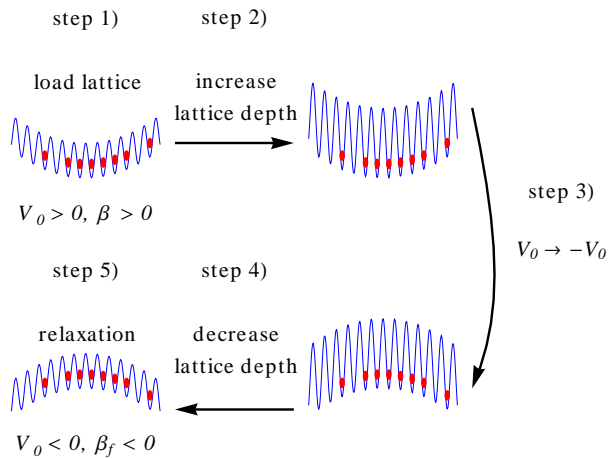}
 \caption{(Color online) \emph{Reversing} the external harmonic potential, to achieve $T<0$ with an ultracold atomic cloud in a deep optical lattice.}
 \label{fig:protocol}
\end{figure}

Although establishing the correspondence between the Hamiltonians $H$ and $-H$ at $\beta > 0$ and $\beta < 0$ is completely trivial theoretically, it represents a major challenge in an experiment since ultracold atomic clouds are always prepared at $T>0$. In order to achieve negative $T$, one has to follow a certain protocol, as discussed in Refs.~\cite{Mosk,negT}, which we review here in the case of fermions in a deep optical lattice.

From the theoretical aspect, the term which determines the sign of $\beta$ is the external potential characterized by $V_0$, since the kinetic and interaction terms (described by the Hubbard model; see the review by Bloch {\textit et al.}~\cite{coldatom-revmod}) are bounded. Therefore the main step is to reverse the external potential $V_0 \to -V_0$ and then let the cloud relax. The reason why the cloud cannot explode is energy conservation: the potential energy of the atoms cannot be reduced indefinitely as it can only be converted to kinetic and interaction energy of the atoms. Furthermore, since the external potential is not bounded from below, equilibrium at positive temperatures is excluded. 

From the experimental side, the protocol involves the following steps (see also Fig.~\ref{fig:protocol}): (1)~Load the cloud to an optical lattice in an (insulating) state at $V_0 >0$ and $\beta >0$; (2)~increase the height of the optical lattice suddenly to ``freeze'' the density distributions; (3)~reverse the sign of the external potential $V_0 \to -V_0$ (and possibly adjust other control parameters, as well); (4)~suddenly reduce the height of the optical lattice; and (5)~wait for the cloud to relax. According to the Boltzmann simulation in Ref.~\cite{negT}, the cloud should be close to an equilibrium state with $T < 0$  within the typical experimental time scale. The protocol works optimally when the initial state is an insulator, somewhat restricting the final parameters. To access a more extended region of the phase diagram, one should change the couplings quasi-adiabatically after step 5.

Although the dynamics restricted by the principle of energy conservation can lead to $T < 0$ in a reversed parabolic potential, the final state could be too hot: the interesting correlated phases typically found in the low-temperature regime might not be accessed experimentally. The main source of heating is that the protocol is inescapably non-adiabatic. Therefore it is important to discuss which are the suitable conditions and estimate how much entropy is generated. The optimal parameters depend on the details of the specific system. Therefore we focus on two, rather different but theoretically important models to show how simulations could benefit from applying $T < 0$.

\section{The attractive SU(3) Hubbard model}

Fermions hopping between nearest-neighbor lattice sites with three possible internal states (colors) and interacting locally with the same interaction strength are described by the SU(3) Hubbard model:
\begin{equation}
H^{SU(3)} = - t_F \sum_{\langle ij\rangle \alpha} [\hat c_{i\alpha}^\dagger \hat c_{j\alpha} + {\rm H.c.}] + U \sum_{i \alpha<\beta} \hat n_{i \alpha}\hat n_{i \beta} ,\label{eq:def:SU3}
\end{equation}
where $t_F > 0$ is the nearest-neighbor hopping amplitude, $U$ is the strength of the local interaction, and $\hat c_{i\alpha}^\dagger$ is a fermionic creation operator at site $i$ for the fermion color $\alpha=1,2,3$. 

\begin{figure}
 \centering
 \includegraphics[width=0.35\textwidth,clip=true]{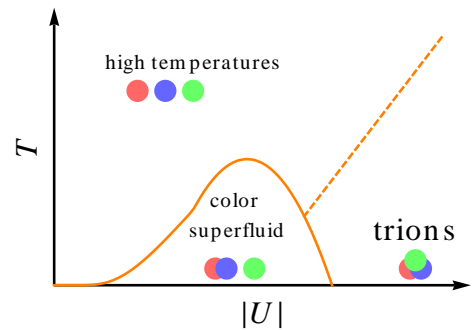}
 \caption{(Color online) Schematic phase diagram of the homogeneous attractive SU(3) Hubbard model. The quantum phase transition between the color superfluid and the trionic phase happens when the interaction $|U|$ is of the order of the bandwidth. }
 \label{fig:su3phasediag}
\end{figure}

The model in Eq.~(\ref{eq:def:SU3}) with an attractive interaction, $U<0$, received considerable attention~\cite{su3attr,su3paper2,su3paper3,su3paper4}: This Hamiltonian, in addition to being a generalization of the two-component case with SU(2) symmetry, also shows similarities to quantum chromodynamics, the gauge theory of quarks describing the strong interaction. In particular, both models exhibit a quantum phase transition between a superfluid phase with broken color SU(3) symmetry (color superfluid) and a phase where three particles form color-singlet bound states (baryons or trions). If we could simulate the attractive SU(3) Hubbard model, we could glimpse some puzzles of this quantum field theory \cite{QCD}. A schematic phase diagram is shown in Fig.~\ref{fig:su3phasediag}.

To realize the model in Eq.~(\ref{eq:def:SU3}), the attention was first focused on $^6$Li atoms in a strong magnetic field, due to the hyperfine structure and the large negative background scattering length $a_{bg}\approx -100$nm\cite{Feshbach-Li6}. Unfortunately, there are certain problems with this isotope. The first one is that the SU(3) symmetry is only approximate as the scattering lengths are different \cite{Feshbach-Li6}. The more substantial problem, however, is that the three-body loss rate was found to be rather large in experiments \cite{3body-Li}. This loss mechanism is not included in the model in Eq.~(\ref{eq:def:SU3}). Although there are approaches where strong three-body loss is used as a dynamical three-body constraint \cite{3body-constraints}, a very important aspect, ``baryon formation'', cannot be studied that way.

It is therefore desirable to find an isotope where three-body loss is much weaker. A possibility is to use $^{173}$Yb, which was already trapped in an optical lattice \cite{Yb-exp}. Actually, there is a simple argument why three-body recombination should be less relevant than in $^6$Li. In cases when the scattering length is much larger than the interatomic van der Waals length, the three-body recombination rate $K_3$ scales with some average scattering length $\bar a_s$ and atomic mass $m$,~\cite{3body-boson,3body-fermion} 
\begin{equation}
 K_3 \sim \frac{1}{m} \bar{a}_{s}^4 .\label{eq:K3}
\end{equation}
As $^6$Li is much lighter and has a larger scattering length than $^{173}$Yb, the loss rates in $^{173}$Yb are expected to be much weaker than in $^6$Li.

There is a further advantage of using the isotope $^{173}$Yb: due to its closed electronic subshell, the state of the atom is determined by the nuclear-spin state. Since the interaction between the atoms is mediated by dipolar van der Waals forces, it is in turn independent of the nuclear spin with a very good approximation. As a consequence, the interactions between $^{173}$Yb atoms will have an increased symmetry, SU(2) $\to$ SU(N), at ultracold temperatures \cite{sun-naturephys}. In Ref.~\cite{Yb-exp}, SU(6) symmetry was established; however, any number $N \leq 6$ of the nuclear states can be populated using optical pumping. The number $N$ is fixed by the initial preparation due to the conservation of atoms in each component. From now on we shall concentrate on the case $N=3$.

The properties of $^{173}$Yb make it a promising candidate to realize the SU(3) Hubbard model; however, the scattering length is positive, $a_{s} \approx 10$nm,~\cite{Yb-exp} leading to \emph{repulsive} interactions as $U \sim a_s$. This is precisely a situation when negative absolute temperatures would allow one to realize a system with \emph{effectively attracting} interactions.

Now we review some properties of the color superfluid and trionic phases based on Ref.~\cite{su3paper3}, where the half-filled SU(3) Hubbard model was studied at finite temperature. The entropy of the color superfluid can be seen in Fig.~2 of Ref.~\cite{su3paper3}. To enter this phase one requires an entropy per site $S/L^3 \approx 0.3$ corresponding to an entropy per particle (with 3/2 atoms per site at half band-filling) $s \approx 0.2$, at a temperature $T \approx 0.2 W$, where $W$ corresponds to the full bandwidth of the non-interacting band. Although the calculations were performed on a Bethe lattice with infinite coordination number, $z \to \infty$, the non-interacting density of states approximates the density of states for nearest-neighbor hopping in a simple cubic lattice. Thus the values listed here should approximately be valid for a $d=3$ fermionic cloud in a cubic optical lattice. Regarding the trionic phase, an artifact of the analysis on the Bethe lattice with $z \to \infty$ is that the trions are localized as their hopping scales with $z^{-3/2}$. Nevertheless, trions in $d=3$ dimensions should compose some Fermi liquid which crosses over to the high-temperature phase of a three-component Fermi liquid.

Unfortunately, the lowest value of the entropy per particle reported for (two-component) fermions is $s_0 \approx \ln 2$ in the center of a trap (with $S_{tot}/N_{tot} \approx 1.3$ for the total cloud)\cite{highTvsDMFT,DeMarco}. The main reason for these high entropies, i.e., hot fermionic clouds, is that Pauli blocking suppresses cooling efficiency at ultracold temperatures. New cooling procedures are being developed to reach substantially lower entropies, e.g., to realize the Neel antiferromagnetic state in the two-component Hubbard model \cite{DeMarco}. In the following, however, we shall consider a ``hot'' fermionic cloud and expand the free energy in the parameter $\beta t_F$ \cite{highTexp}. A reasonable agreement between high-temperature expansions and dynamical mean-field theory has been found in Ref.~\cite{highTvsDMFT} for typical experimental parameters.

We consider the free energy $F=F(\beta,\mu,U,W)$ of the fermions, defined by
\begin{equation}
 -\beta F = \ln {\rm Tr} \exp [- \beta ( H^{SU(3)} - \mu \hat N + W \sum\limits_i \hat t_{i}) ],
\end{equation}
for a given chemical potential $\mu$. Here we included also a three-body term $\sim W$ to be able to determine the density of triple occupancies $\hat t_i \equiv \hat n_{i1} \hat n_{i2} \hat n_{i3} $. We find that for a simple cubic lattice, the free-energy density $f=F/L^3$ can be expanded to second order in $\beta t_F$ as~\cite{highTexp}
\begin{eqnarray}
 -\beta f &=& \ln z_0 + \frac{3 (\beta t_F)^2}{z_0^2} X_2 + {\cal O}(\beta t_F)^4 , \label{eq:FhighT}
\end{eqnarray}
with
\begin{equation}
z_0 = z_0(\beta,\mu,U,W) = 1+3 \zeta+3 \zeta^2 u + \zeta^3 u^3 w
\end{equation}
and
\begin{eqnarray}
 X_2 = X_2(\beta,\mu,U,W) &=& \phantom{+} 3 \zeta + 12 \zeta^2 \frac{1- u}{\beta  U}  \nonumber \\
	&& + 6 \zeta^3 u \left(2  +\frac{ 1 - u^2 w }{\beta (2 U+W)} \right) \nonumber \\
	&& +12 \zeta^4 u^2 \frac{1 - u w}{\beta (U+W)} +3 \zeta^5 u^4 w, \label{eq:su3-X2}
\end{eqnarray}
with $\zeta = e^{\beta \mu}$, $u=e^{-\beta U}$, and $w=e^{-\beta W}$.
The relevant local quantities are calculated analytically in local-density approximation as derivatives of the free energy according to the usual definitions,
$n=-\partial f/\partial \mu$, $d=\partial f/\partial U$, $t=\partial f/\partial W$, $k=-\partial f/\partial t_F$, and $s=-\partial f/\partial T$, taken at $\mu = \mu_0-V_0 {\mathbf r}^2$ and $W=0$.
These correspond to the local densities of total particle number, total double occupancy, triple occupancy, hopping amplitudes, and entropy, respectively. 

Now we turn to the protocol shown in Fig.~\ref{fig:protocol} to reach the state with effectively attracting interactions at $T<0$. The high-temperature expansion describes both the initial and the final states of the cloud in an optical lattice, since it can be applied for both $\beta >0$ and $\beta < 0$. Initially the system is in equilibrium with parameter values $V_0 = 0.01 t_F$, $U=t_F$, and $\beta = 0.5/t_F$. We varied the central chemical potential $\mu_0$ so the total particle number $N_{tot}$ is changed. For the parameters being used, $N_{tot} \approx 5\times 10^4$ corresponds to a metallic cloud with an entropy $S_{tot}/N_{tot}\approx 2.4$, while $N_{tot} \approx 6.5\times 10^5$ corresponds to a band insulator with $S_{tot}/N_{tot}\approx 0.75$. Thus realistic values of the entropy per particle are covered by the calculations.

The final state corresponding to $\beta_f < 0$ is determined by energy (and particle) conservation. We approximate the density distributions at step 4 with the initial equilibrium distributions. Therefore we have to solve
\begin{eqnarray}
  \int\!\!d^3 \mathbf{r} \left[- t_F (1-q) k_i(\bfr) + U d_i(\bfr) - V_0 r^2 n_i(\bfr) \right] \nonumber\\ 
	=\int\!\!d^3 \mathbf{r} \left[ -t_F k_f(\bfr) + U d_f(\bfr) - V_0 r^2 n_f(\bfr) \right]; \nonumber \\
  \int\!\!d^3 \mathbf{r} \, n_i(\bfr) = \int\!\!d^3 \mathbf{r} \, n_f(\bfr),
\end{eqnarray}
for the final inverse temperature, $\beta_f$, and the final central chemical potential, $\mu_{0f}$. Here we used abbreviations for
initial and final quantities, e.g., the particle density is $n_i(\bfr) = n(\bfr;\beta,\mu_0,V_0,U)$ and $n_f(\bfr) = n(\bfr;\beta_f,\mu_{0f},-V_0,U)$, respectively. The quantity $q$ corresponds to the relative dephasing of nearest-neighbor sites during step 3 \cite{negT}: while the external potential is reversed, the cloud is held for some waiting time $\tau_w$ in a deepened optical lattice with a quenched hopping $t_F \to 0$. During this step, each site acquires a different local phase due to the inhomogeneity induced by the external potential. When the optical lattice is finally relaxed during step 4, the kinetic energy represents a sum of terms with random complex phases, which averages to zero if $\tau_w$ is large enough. In our calculations, we used two cases of no ($q=0$) or complete ($q=100\%$) dephasing.

\begin{figure}
 \centering
 \includegraphics[width=0.42\textwidth,clip=true]{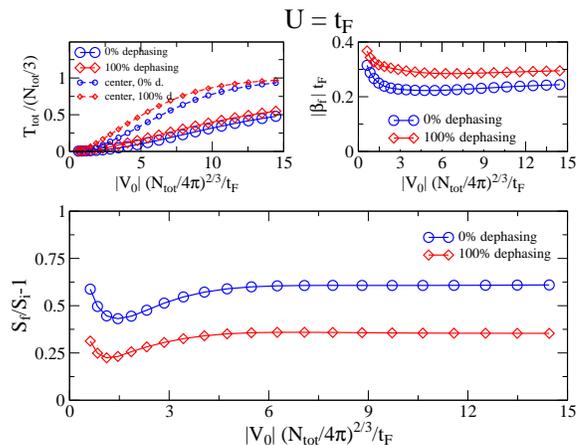}
 \caption{ (Color online) Lower panel: The relative change in the entropy as a function of the compression, after reversing $V_0 \to -V_0$ in the SU(3) Hubbard model for $U= t_F$. Upper right panel: the final inverse temperature $|\beta_f|$. Upper left panel: Relative total trion occupation, $T_{tot}/(N_{tot}/3)$ in the final state is shown by the large symbols, while the relative trion occupation in the center, $t(\bfr =0)/(n(\bfr = 0)/3)$ is shown by the small symbols.  \label{fig:su3}}
\end{figure}

After obtaining the parameters $\beta_f$ and $\mu_{0f}$ numerically, we can calculate all quantities in the final state.
We show the relative entropy change, $S_{\rm f}/S_{\rm i}-1$, the final inverse temperature $|\beta_f|$, and the relative ``trion content'' $T_{tot}/(N_{tot}/3)$ as a function of the compression $|V_0| \left(\frac{N_{tot}}{4\pi}\right)^{2/3}$ on Fig.~\ref{fig:su3}. (For an inhomogeneous system in a parabolic potential, the compression describes the system in the thermodynamic limit, where $N_{tot}\to\infty, V_0 \to 0$, while $V_0 N_{tot}^{2/d}$ is fixed \cite{SU2HUbbard}.) 
We see that deep in the band insulating regime, the entropy increase is around 30 \% (or 60\%) in the case of complete (or no) dephasing. These values are not too high in comparison to an $\approx$100\% increase in the entropy of a two-component Fermi gas when it is first loaded in an optical lattice and then unloaded from it \cite{highTvsDMFT}. The final values of the entropy with the high trionic occupations should allow us to explore the crossover from the high-temperature phase to the trionic phase. In order to reach the color superfluid phase, however, additional reduction of the entropy is necessary.

\section{S=1 antiferromagnetic spin chain}

\begin{figure}
 \centering
\includegraphics[width=0.3\textwidth,clip=true]{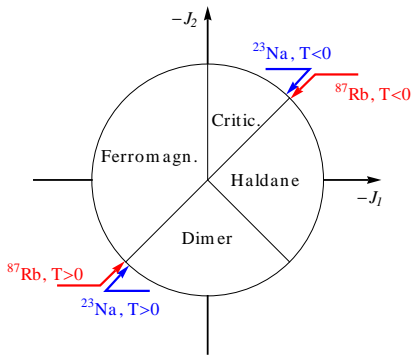}
\caption{(Color online) Schematic phase diagram of the $d=1$ quadratic-biquadratic spin model, after Ref.~\cite{spin1}. Arrows indicate parameters that apply to realizations with $^{87}$Rb and $^{23}$Na isotopes. }
\label{fig:QBphasediag}
\end{figure}

In our second example, we consider spin-1 bosons in a $d=1$ dimensional lattice. Two ultracold bosonic atoms with spins $S_1,S_2=1$ can scatter in total spin sectors $S_1+S_2= 0$ and  $S_1+S_2= 2$. As a consequence, the appropriate Hamiltonian in a deep enough optical lattice in a homogeneous situation is~\cite{PethickSmith,boson-s1}
\begin{eqnarray}
 H^{S1BH} &=& -t_B \sum_{\langle ij\rangle} \sum_{\sigma}[\hat b_{i\sigma}^\dagger \hat b_{j\sigma} + {\rm H.c.}] \label{eq:def:S1BH} \\
	&& + \frac{U_0}{2}\sum_{i} \hat n_i (\hat n_i - 1 )+ \frac{U_2}{2}\sum_{i} ( \hat \mathbf{S}_i^2 - 2\hat n_i) , \nonumber 
\end{eqnarray}
where $t_B >0$ is the hopping amplitude; $\hat b_{i\sigma}$ are bosonic operators with spin projections $\sigma=+1,0,-1$; $\hat n_i = \sum_\sigma \hat b_{i\sigma}^\dagger \hat b_{i\sigma}$ is the number of bosons at a site; $U_0 \sim (a_0 + 2 a_2) $ and $U_2 \sim (a_2 - a_0)$ are the interaction parameters with $a_S$ being the $s$-wave scattering lengths for scattering in the $S=0,2$ spin channels; and $\hat \mathbf{S}_i$ being the spin operator at site $i$. Typically, $a_0 \approx a_2$, therefore $|U_2| \ll |U_0|$. For $t_B \ll U_0 $, a Mott insulating phase is realized when the filling is one boson per lattice site. At low energies in the Mott phase, one can describe the system by the so-called quadratic-biquadratic spin model\cite{boson-s1,spin1,spin1-QZE}
\begin{equation}
 H^{QB} = - J_1 \sum_{\langle ij\rangle} {\mathbf S}_i \cdot {\mathbf S}_j- J_2 \sum_{\langle ij\rangle} ({\mathbf S}_i \cdot {\mathbf S}_j)^2 . \label{eq:def:QB}
\end{equation}
The couplings are given by\cite{boson-s1}
\begin{eqnarray}
 \frac{J_1}{t_B^2} &=& \frac{2}{U_0 +U_2}, \nonumber \\
 \frac{J_2}{t_B^2} &=& \frac{2}{3} \frac{1}{U_0 +U_2}+ \frac{4}{3} \frac{1}{U_0 - 2 U_2} . \label{eq:J1J2}
\end{eqnarray}
The model in Eq.~(\ref{eq:def:QB}) in $d=1$ dimension has a rich ground state phase diagram, shown in Fig.~\ref{fig:QBphasediag} (see also Ref.~\cite{spin1} and references therein; a phase diagram in a magnetic field can be found in Ref.~\cite{spin1-QZE}). In addition to these intriguing phases, even finite temperature ($T \neq 0$) properties are interesting. In the last decades, major theoretical efforts have been made to calculate $T \neq 0$ dynamical correlation functions of (gapped antiferromagnetic) spin chains \cite{AF-spinchains}. A simulation with ultracold atoms could provide an important experimental benchmark. However, since bosons collapse for $U_0 < 0$, so far only the $J_1,J_2 > 0$ regime could be realized. Negative absolute temperatures provide a possibility to reach the $J_1,J_2 < 0$ regime.

To realize the antiferromagnetic couplings, we revisit the approach of Ref.~\cite{spin1} and modify it to fit to the general protocol shown in Fig.~\ref{fig:protocol}. In Ref.~\cite{spin1} constant electric and staggered magnetic fields have been proposed to change the couplings and to reach the ground state of $H$ from the ground state of $-H$. An energy gap is essential for the adiabatic evolution; i.e., the critical phase cannot be accessed. Application of constant electric fields is also not ideal for two reasons. First, there is no equilibrium for $\beta \neq 0$ in the thermodynamic limit in the presence of a constant electric field for a homogeneous system. Second, as discussed in Ref.~\cite{grav-expansion}, a finite cloud in the presence of constant forces may exhibit a behavior that is substantially different from the corresponding homogeneous system. Such dynamics should be avoided when the aim is to simulate a specific quantum system.

Although we think that application of electric fields should be avoided, staggered magnetic fields are necessary for optimal results. To realize this staggered field, one can apply a secondary, spin-dependent optical lattice with a lattice period twice as large as the original which shifts the energies of the $S^z = \pm 1$ atoms in opposite directions~\cite{spin1}. It can be represented by the term
\begin{eqnarray}
 H_{st} &=& -B \sum_i (-1)^i \, {\mathbf S}_i^z  \\
	&=&  -B\!\!\!\!\!\sum_{i,\sigma=+1,0,-1}\!\!\!\!\!(-1)^i \,\sigma\, \hat b_{i\sigma}^\dagger \hat b_{i\sigma} . \nonumber
\end{eqnarray}
The improved protocol to reach $T<0$ with spin-1 bosons involves taking both $V_0 \to - V_0$ and $B \to -B$ during step 3. This latter could be implemented experimentally by shifting the phase of the lasers that create the secondary optical lattice by $\pi$. Note that in contrast to the case of spinless bosons, discussed in Ref.~\cite{negT}, the interaction $U_0>0$ is not to be changed. This repulsive interaction translates to \emph{effectively} attracting bosons at $\beta < 0 $, and to antiferromagnetic couplings $J_1, J_2 <0$ in Eqns.~(\ref{eq:J1J2}).

We would like to emphasize that, in contrast to the instability of a spinor Bose gas prepared with $a_S <0$, suddenly changing from the Mott insulator $U_0 \gg t_B$ phase to the effectively attractive case $\tilde U_0 = -U_0 < 0$ is metastable. Since only the atoms themselves can transport energy, the binding energy of two bosons has to be taken away by a large number $\approx U_0/t_B$ of other atoms. Nevertheless, such a configuration has a very low probability. As a consequence, double occupancies can form only at an exponentially suppressed rate. A similar discussion was applied for a complementary situation in Ref.~\cite{doublondecay}.

Relaxation to an equilibrium state in a $d=1$ dimensional system is a delicate and extremely difficult problem, especially in the vicinity of points where the model is integrable \cite{closed-q-dynamics}. In fact, this is an additional reason why the experimental simulation of the spin model in Eq.~(\ref{eq:def:QB}) is important.
To resolve the problem of one-dimensional thermalization and to simulate the conditions in experiments, we consider an \emph{array} of
chains instead of a single chain: The bosons are most likely trapped initially in an anisotropic $d=3$ dimensional optical lattice, where hopping
along the chains, characterized by $t_B$, is stronger than hopping between the chains, given by $t_B^\perp$. Even weak interchain couplings should break integrability and allow relaxation to a thermal state. Note that with orthogonally arranged laser beams, $t_B$ and $t_B^\perp$ can be changed independently. Although one could also tune the aspect ratio of the parabolic potential, we focus on a spherically symmetric external potential.

For the remainder of this section we focus on a cloud with interaction strengths $U_0 = 50 t_B, U_2 = -0.01 U_0$, but similar results can be found for $ U_2 = 0.04 U_0$. These values apply for $^{87}$Rb and $^{23}$Na, respectively \cite{spinor-Ho}. Initially, the external potential is set to $V_0 = 0.005 t_B$, and the central chemical potential $\mu_0$ is changed to vary the number of particles $N_{tot}$ and, consequently, the compression. The value of the interchain hopping is $t_B^\perp = 0.1 t_B$.

To describe the initial system, we use a Gutzwiller wave function of the form
\begin{eqnarray}
 \vert G \rangle &=& \prod_i \left[ f_0(i)+ \sum_\alpha f_\alpha(i) \hat a_{i\alpha}^\dagger +  \sum_\alpha \frac{f_{\alpha\alpha}(i)}{\sqrt{2}} (\hat a_{i\alpha}^\dagger )^2 \right. \nonumber \\
	&&  \phantom{\prod}+ \left. \sum_{\alpha<\beta}   f_{\alpha\beta}(i) \hat a_{i\alpha}^\dagger \hat a_{i\beta}^\dagger \right] \vert 0 \rangle \label{eq:def:G}
\end{eqnarray}
where $f_\lambda(i)$ correspond to occupation amplitudes of local configurations, and the boson operators $\hat a_{i\alpha}^\dagger$ are related to the operators $\hat b_{i\sigma}^\dagger$ in Eq.~(\ref{eq:def:S1BH}) by a canonical transformation,~\cite{boson-s1}
\begin{equation}
 a_z^\dagger = b_0^\dagger, a_x^\dagger = \frac{b_{-1}^\dagger - b_{+1}^\dagger }{\sqrt{2}}, a_y^\dagger =i \frac{b_{-1}^\dagger + b_{+1}^\dagger }{\sqrt{2}} .
\end{equation}
In Eq.~(\ref{eq:def:G}) we neglected triple- and higher occupancies which are almost completely suppressed in the strongly repulsing regime with at most one boson per lattice site.
The Gutzwiller wave function works in the Mott insulating region well, while it is expected to shift the boundary of the Mott region
in comparison to the exact value \cite{coldatom-revmod}. Since we focus on a tightly compressed cloud ($\mu_0 \approx U_0 \gg t_B$) in a relatively strong staggered field $B > J_1,J_2$, the shifts of values of the total energies and total atom numbers (the quantities relevant for our calculations) are expected to be small.

The variational energy per lattice site can be expressed as (with $\langle G \vert G \rangle=1$)
\begin{eqnarray}
 E_v &\equiv& \langle G \vert H^{S1BH} + H_{st} - \mu \sum_{i\alpha} \hat n_{i\alpha}  \vert G \rangle/L^3 \nonumber \\
	&=& -\phantom{2}t_B \sum_{\alpha} [Q_\alpha^*(A)Q_\alpha(B)+Q_\alpha^*(B)Q_\alpha(A)]\nonumber \\
	&& -2t_B^\perp \sum_{\alpha} [Q_\alpha^*(A)Q_\alpha(B)+Q_\alpha^*(B)Q_\alpha(A)] \nonumber \\ 
	&& - \mu \sum_\alpha \frac{n_\alpha(A) + n_\alpha(B) }{2}  - B \frac{m_A-m_B}{2} \nonumber \\
	&&+ \frac{U_0}{2} \frac{d_A+d_B}{2} + \frac{U_2}{2} \frac{\tilde d_A+\tilde d_B}{2} \label{eq:2site-Gutzwiller}
\end{eqnarray}
with
\begin{eqnarray}
 Q_\alpha(l) &=& f_\alpha^*(l) f_0(l) + \sqrt{2} f_{\alpha\alpha}^*(l)f_\alpha(l) + \sum_{\beta \neq \alpha} f_{\alpha\beta}^*(l)f_\beta(l), \nonumber \\
 n_\alpha(l) &=& |f_\alpha(l)|^2 + 2 |f_{\alpha\alpha}(l)|^2 +\sum_{\beta \neq \alpha} |f_{\alpha\beta}(l)|^2, \nonumber \\
 m_l\phantom{(l)} &=& i \Big\lbrace f_y^*(l) f_x(l) +  f_{yz}^*(l)f_{xz}(l) \nonumber \\
	&& + \left.  \sqrt{2} f_{xy}^*(l) [f_{xx}(l)-f_{yy}(l)] - {\rm H.c.} \right\rbrace \nonumber \\
 d_l\phantom{(l)} &=& 2\sum_\alpha |f_{\alpha\alpha}(l)|^2 + 2 \sum_{\beta < \alpha} |f_{\alpha\beta}(l)|^2 \nonumber \\
 \tilde d_l\phantom{(l)} &=& -2 \sum_{\alpha \neq \beta} f_{\alpha\alpha}^*(l) f_{\beta\beta}(l) + 2 \sum_{\beta < \alpha} |f_{\alpha\beta}(l)|^2 \;,
\end{eqnarray}
where $l=A,B$ denotes sites in the two sublattices. One can reduce the number of variational parameters by using the sublattice symmetry.

For given values of the control parameters one can minimize Eq.~(\ref{eq:2site-Gutzwiller}) with respect to the variational parameters $f_\lambda$. It is confirmed that double occupancies are negligible for the large value $U_0 = 50 t_B$. The initial energy is calculated using the densities obtained by the Gutzwiller ansatz in local-density approximation, e.g., $n_\alpha(\bfr) = n_\alpha(\mu=\mu_0-V_0 \bfr^2)$. This gives
\begin{eqnarray}
 && E_i(U_0,U_2,V_{0f}=-V_0,B_f=-B) \label{eq:Einitial-spin1} \\
 &&= \int\!\!d^3\bfr \left\lbrace -t_B(1+2\kappa)(1-q)\!\sum_\alpha [ Q_{A\alpha}^{*}(\bfr) Q_{B\alpha}(\bfr)+ {\rm h.c.}] \right. \nonumber \\
 && \phantom{mmmm} + V_{0f} \bfr^2 \sum_\alpha [n_{A\alpha}(\bfr)+n_{B\alpha}({\bfr})]/2 \nonumber \\
 && \phantom{mmmm} - B_f [m_A({\bfr}) - m_B({\bfr})]/2 \nonumber \\
 && \phantom{mmmm} \left. +\frac{U_0}{4} [d_A({\bfr})+d_B({\bfr})] + \frac{U_2}{4} [\tilde d_A({\bfr})+\tilde d_B({\bfr})] \right\rbrace \nonumber
\end{eqnarray}
Here $\kappa = t_B^\perp/t_B = 0.1$. Similarly to the fermionic case, the coefficient $q$ corresponds to the dephasing efficiency during step 3.

\begin{figure}
 \centering
 \includegraphics[width=0.4\textwidth,clip=true]{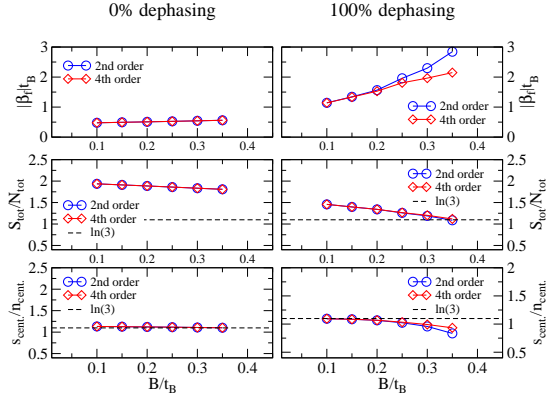}
 \caption{(Color online) Final parameters for the S=1 bosonic Hubbard model in a staggered field at a compression of $|V_0| (3 N_{tot}/4\pi)^{2/3} /t_B = 12$ for $t_B^\perp = 0.1 t_B$, $U_0 = 50 t_B$ and $U_2 = -0.01 U_0$. The final entropies and final inverse temperatures are shown for the cases of $q=0\%$ and $q=100\%$ dephasing of the initial kinetic energy. The different lines correspond to results calculated in 2nd and 4th-order expansions in $\beta t_B$.}
 \label{fig:staggared}
\end{figure}

\begin{figure}
 \centering
\includegraphics[width=0.4\textwidth, clip=true, trim = 0 0 0 2cm]{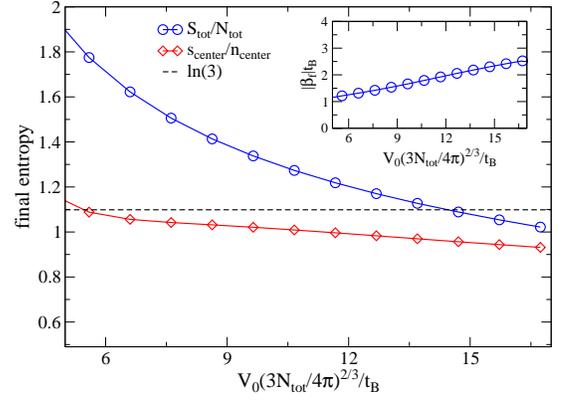}
 \caption{\label{fig:s1boson-B=0.3} (Color online) The final entropy of spin-1 bosons as a function of the compression for a staggered field value $B = 0.3 t_B$ with 100\% dephasing of the initial kinetic energy, using fourth order expansion. Inset: the final inverse temperature $\beta_f$. Values of $t_B^\perp$, $U_0$ and $U_2$ are the same as in Fig.~\ref{fig:staggared}.}
\end{figure}

Let us now consider the final state. We assume that a high-temperature expansion\cite{highTexp} in $\beta t_B$ can be applied. The validity of this step has to be confirmed later. As we have reasoned above, all processes where a doubly occupied site is created are exponentially suppressed and will not occur during the typical time scale of an experiment. We therefore assume that the system relaxes in a subspace where two-body occupations are absent. Therefore we use the following expansion of the free energy:
\begin{eqnarray}
 -\beta f^{S1BH} &=& \ln z_0 + \frac{(\beta t_B)^2  + 2 (\beta t_B^\perp)^2 }{z_0^2}  X_2  \nonumber \\
	&& + (\beta t_B)^4 \left[ \frac{1}{z_0^2} X_{4a} + \frac{1}{z_0^3} X_{4b} - \frac{3}{2} \frac{1}{z_0^4} X_2^2 \right] \nonumber \\
	&& + {\cal O}{\mathbf (}(\beta t_B)^6 , \beta^4 t_B^2 (t_B^\perp)^2 {\mathbf )}
\end{eqnarray}
with
\begin{equation}
z_0 = 1+\zeta (1+2 {\cal B} ),
\end{equation}
\begin{equation}
X_2 = \zeta \left(1+2  {\tilde {\cal B}}  \right),
\end{equation}
\begin{eqnarray}
X_{4a} = \frac{\zeta}{12}\left[1+\frac{6}{\beta^2  B^2} ( { {\cal B}}  - {\tilde {\cal B}}   ) \right],
\end{eqnarray}
and
\begin{eqnarray}
 X_{4b} &=& \phantom{+} \frac{\zeta}{6}\left[1+\frac{6}{\beta^2  B^2} ( { {\cal B}}  - {\tilde {\cal B}}   )\right] \nonumber\\
	&& + \frac{\zeta^2}{6}\left[ 1 - \frac{6}{\beta^2  B^2} \left({\cal B} -(1 +\beta^2  B^2 )  {\tilde {\cal B}} \right)\right. \nonumber \\
	&& + \left. \frac{3}{\beta^2  B^2} ({\cal B}^2 + \beta^2 B^2  {\tilde {\cal B}}^2 - { {\cal B}}  {\tilde {\cal B}} )   \right].  
\end{eqnarray}
where we introduced $\zeta=e^{\beta\mu}$, ${\cal B} = \cosh(\beta B)$, and $ {\tilde {\cal B}} = \sinh(\beta B)/(\beta B)$.
The densities of different quantities in local-density approximation $[ \mu(\mathbf r)=\mu_0-V_0 {\mathbf r}^2 ]$ can be calculated analytically from definitions like
$m_s = - \partial f/ \partial B$, $ k_{||} = -\partial f/\partial t_B$, and  $k_{\perp}=-\partial f/\partial t_B^\perp$.
Finally, the energy in the final state is given by 
\begin{eqnarray}
 E_f &=& \int\!\!{d^3\bfr}\,\left[ -\phantom{2}t_B k_{||}(\beta_f,\mu_{0f},B_f,{\bfr}) \right.  \\
	&& \phantom{nnnni} -2t_B^\perp k_{\perp}(\beta_f,\mu_{0f},B_f,\bfr)  \nonumber \\
	&& \left. - B_f m_s(\beta_f,\mu_{0f},B_f,{\bfr}) +V_{0f}{\bfr}^2 n(\beta,\mu_0,B_f,{\bfr}) \right],\nonumber
\end{eqnarray}
where $B_f=-B$ and $V_{0f}=-V_0$. Total-energy conservation implies that $E_f$ equals the initial energy $E_i$ given by Eq.~(\ref{eq:Einitial-spin1}). Considering also the conservation of particles $N_f = N_i$, one can find numerically the final inverse temperature $\beta_f <0$ and the final central chemical potential $\mu_{0f}$, which determine the final state. In Fig.~\ref{fig:staggared} we show the final temperatures and the final entropies as a function of the staggered magnetic field for a given compression $|V_0| (3 N_{tot}/4\pi)^{2/3} /t_B = 12$, while in Fig.~\ref{fig:s1boson-B=0.3} we show quantities for $B/t_B=0.3$ as a function of the compression. As a check of the validity of the approach, we compare the results using the second- and fourth-order expansions.

As we can see, for low values of $B$, and especially with no dephasing ($q=0$) during step 3, the entropy per particle is large, $S_{tot}/N_{tot} \geq \ln 3$. This characteristic value corresponds to a system where sites are populated by precisely one of three kind of particles in an uncorrelated fashion. Higher entropies indicate considerable fluctuations in not just the spin, but also in the occupation of the sites. This implies that the spin Hamiltonian defined in Eq.~(\ref{eq:def:QB}) should not be used to describe the model, and that the application of the staggered magnetic field is necessary. As the strength of the staggered field is increased, both the entropy and $|T_f|$ decrease. However, the results of second- and fourth-order expansions start to deviate significantly for $B/t_B \geq 0.35$. Further increasing $B$ eventually leads to the breakdown of the high-temperature expansion. Although we cannot give quantitative estimates based on our method, stronger staggered magnetic fields should lead to even lower entropies based on a simple physical picture. As $H_{st}$ becomes the dominant term in the effective Hamiltonian, the initial densities at $\beta \to +\infty$ and $B>0$ approximate the densities in the final state, with $\beta_f <0$ and $B_f = -B$, very well in the Mott insulating center. The heating occurs mainly at the compressible edges. It is important, however, that $B \ll U_0$ is maintained for metastability as the local Zeeman energy of two atoms has to be much smaller than the interaction energy. After step 5, $B_f$ and $t_B^\perp$ have to be weakened adiabatically to reach the Hamiltonian in Eq.~(\ref{eq:def:QB}) with antiferromagnetic couplings. We emphasize that exploring finite temperature dynamics of spin chains experimentally would be as interesting as establishing the ground state phase diagram.

\section{Conclusions}

We have discussed two important theoretical models which can be simulated with an ultracold atomic cloud at \emph{negative absolute temperatures}. In the first case, repulsively interacting fermionic atoms are considered to realize the attractive SU(3) Hubbard model, while in the second case, bosonic atoms with spin $S=1$ in the Mott phase at $T < 0$ could simulate antiferromagnetically coupled spin chains. In general, considering negative temperatures provides an alternative way to realize couplings which are hard to reach at $T>0$.

We would like to emphasize that the models discussed here could be simulated in other ways. For example, in the case of the bosons one could reverse the interaction $U_0 \to -U_0$ using Feshbach resonances during step 3 instead of reversing the external potential $V_0$.

It would be interesting to study the real-time dynamics of the relaxation in the two models. In the case of the fermionic cloud, we expect that energy diffusion determines the time scales, similarly to the two-component case \cite{negT,grav-expansion}. The relaxation in the bosonic case is, however, different and much more challenging.

\acknowledgments
I thank Luis Santos and especially Achim Rosch for discussions and for reading the manuscript. This research has been supported by the SFB 608 and the SFB TR 12 of the Deutsche Forschungsgemeinschaft and by the excellence cluster QUEST.

\end{document}